# On the mechanical properties of novamene: a fully atomistic molecular dynamics and DFT investigation

Eliezer Fernando Oliveira[1,2], Pedro Alves da Silva Autreto[1,3], Cristiano Francisco Woellner[1,2] and Douglas Soares Galvao[1,2*]

[1]Group of Organic Solids and New Materials (GSONM), Gleb Wataghin Institute of Physics, University of Campinas (UNICAMP), Campinas, SP, Brazil

[2]Center for Computational Engineering & Sciences (CCES), University of Campinas - UNICAMP, Campinas, SP, Brazil

[3]Center of Natural Human Science, Federal University of ABC (UFABC), Santo Andre, SP, Brazil

*Corresponding author: galvao@ifi.unicamp.br

**ABSTRACT**

We have investigated through fully atomistic reactive molecular dynamics and DFT simulations, the mechanical properties and fracture dynamics of novamene, a new 3D carbon allotrope structure recently proposed. Our results showed that novamene is an anisotropic structure with relation to tensile deformation. Although novamente shares some mechanical features with other carbon allotropes, it also exhibits distinct ones, such as, extensive structural reconstructions (self-healing effect). Novamene presents ultimate strength (~ 100 GPa) values lower than other carbon allotropes, but it has the highest ultimate strain along the z-direction (~ 22.5%). Although the Young's modulus (~ 600 GPa) and ultimate strength values are smaller than for other carbon allotropes, they still outperform other materials, such as for example silicon, steel or titanium alloys. With relation to the fracture dynamics, novamene is again anisotropic with the fracture/crack propagation originating from deformed heptagons and pentagons for x and y directions and broken $sp^3$ bonds connecting structural planes. Another interesting feature is the formation of multiple and long carbon linear chains in the final fracture stages

# INTRODUCTION

During the last decades, carbon-based structures has been one of the active areas in material science research. This is due in part to the versatility of carbon atom hybridization (sp, $sp^2$, and $sp^3$), which can produce a plethora of possible materials with equal diversity on mechanical, thermal and electronic properties [1,2].

Using only one hybridization type, it is possible to produce allotropes with different topologies and dimensionalities, such as fullerenes (0D), carbon nanotubes (1D), graphene (2D), graphite (3D), and diamond (3D). The mix of different hybridizations in the same structure increases this number of possibilities substantially, and using this approach it is possible to generate 2D and 3D new allotrope structures [3-9].

Some of these proposed structures are based on mixing $sp^2$ regions (graphene-like) with $sp^3$ ones (diamond-like) in the same material. One example is a series of ultrastrong (two times than usual ceramics), hard and exceptionally lightweight carbons recently produced by compressing $sp^2$ hybridized glassy carbon at several temperatures [10]. These structures are formed by a local buckling in graphene membranes using $sp^3$ nodes, which results in an interpenetrating graphene network with long-range order. Those structures have also exceptional compressive strengths with elastic recovery after local deformations. This type of carbon-based structure is an optimal ultralight, ultra strong material for a wide range of multifunctional applications, and the synthesis methodology demonstrates potential to access entirely new metastable materials with exceptional properties [11].

Another example within these proposed 3D carbon allotropes, is the family of the so-called novamene structures [6]. They were recently proposed by L. A. Burchfield *et al.* [6]. They have a 3D structure composed of a combination of hexagonal diamond [12, 13] ($sp^3$ hybridization called lonsdaleite) and hexagonal carbon rings ($sp^2$ hybridization) [14,15] (see Figure 1). The number of hexagonal carbon rings determines the structure type in the family, such as, single-ringed novamene, double-ringed novamene, triple-ringed novamene, and so on [6]. From a topological point of view, novamene can be also considered as a hexagonal diamond structure ($sp^3$) "doped" by carbon hexagons ($sp^2$). First-principles simulations have shown that the single-ringed novamene (Figure 1(a)), presents an electronic indirect bandgap of ~0.3 eV and could be an excellent candidate to be used in electronic devices, such as integrated circuits, optoelectronics, Hall effect sensors, among others [1,2,6]. These complex topologies of mixing of $sp^2$ and $sp^3$ can also result in interesting mechanical properties, but they have not been yet investigated to novamenes and it is one of the objective of the present work.

We investigated the electronic and mechanical properties of single-ringed novamene (hereafter novamente) family member through first principles (DFT) and fully atomistic reactive molecular dynamics calculations. Besides determining the elastic properties, we have also investigated the fracture mechanisms of the structures in tensile/elongation regimes.

**MATERIALS AND METHODS**

We present in Figure 1(a) the geometrical structure of single-ringed novamene, accordingly to the proposed model by L. A. Burchfield *et al.* [6]: a hexagonal crystalline arrangement, with a unit cell size (inset I) of 8.40Åa0 x 8.40Åb0 x 4.99Åc0 (52 carbon atoms). The carbon atoms are colored differently by regions, in order to help to better visualize the material topology (Figure 1a - inset II). The violet colored areas are related to an atomic arrangement formed by $sp^2$ bonds, which are connected to three carbon atom pentagons, in blue. The yellow and the green regions are non-planar hexagons and heptagons, respectively, that are connected by $sp^3$ bonds. The orange areas are related to the external hexagons that form a plane, which are below of the red atoms.

From a tridimensional point of view, novamene can be considered as an ABA'B' packing along the z-direction (Figure 1(b)). The BB' planes are connected in regions without pentagons by a set of carbons (purple in Figure 1(a) - inset III and Figure 1(b)). Such atoms can be "switched" between B and B' planes ($sp^3$ to $sp^2$ hybridization and vice-versa). By "switching" atoms we meant atoms that can alternately connect up and down BB' planes (see details in ref. [6]).

The molecular dynamics (MD) simulations were carried out using the ReaxFF force field [16], as implemented in the computational code Large-scale Atomic/Molecular Massively Parallel Simulator (LAMMPS) [17]. Periodic boundary condition (PBC) were used. The considered supercell size was $5a_0 \times 5b_0 \times 5c_0$ (6500 atoms), which is large enough to avoid spurious size effects (such as, contributions of low energy/long wavelength oscillations) [18]. The initial novamene geometry was energetically minimized using conjugate gradient technique, then followed by a thermal equilibration at room temperature (300K) in an NPT ensemble at the constant hydrostatic pressure of 0 GPa. We observed in this step that novamene structure is stable and the supercell size does not change significantly. After the system thermal equilibration, a uniaxial tensile strain rate of $10^{-6}$A/fs (up to 50%) was applied to deform the structure. The stress-strain behavior was calculated by strain applied along x, y, and z-direction. During the uniaxial tensile procedure, a NPT ensemble was used keeping the system temperature at 300 K and a pressure of 0 GPa for the perpendicular directions of the tensile one.

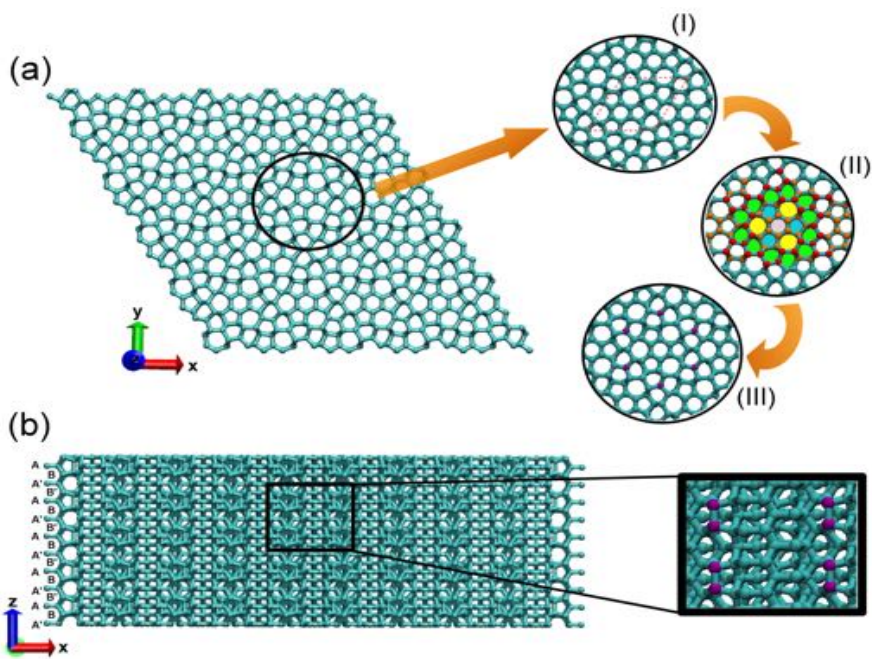

**Figure 1**. Geometrical structure of a single-ringed novamene supercell. (a) Top of view: (I) top representation of a unit cell; (II) orange and red atoms belong to different planes and the arrangement of the carbon atoms form hexagons (in violet), pentagons (in blue), distorted heptagons (in yellow), and distorted hexagons (in green); (III) "switching" atoms that can alternately connect two BB' stacking layers (see part (b)). (b) Lateral view representing the ABA'B' layer stacking; the inset shows how the "switching" atoms connect BB' planes.

We also calculated the second invariant of the deviatoric stress tensor, the von Mises stress [19] for each atom during the straining process. The von Mises yield criterion suggests that fracture starts when this stress value reaches the yield strength of the material [19]. For all MD runs, a timestep of 0.25 fs was considered. This general stress-strain procedure, timestep, and strain rate were already proved to be effective to study the mechanical properties of carbon nanostructures [20].

Due to the large size of the unit cell, the complete strain-stress and fracture analyses using *ab initio* methods is cost-prohibitive, but obtaining the elastic constant and band structure for static calculations is feasible. Also, it is well-known that although fracture dynamics is well described by ReaxFF, it tends to underestimates the elastic constants [20,21], thus the use of *ab initio* methods will provide a better estimate of these values.

Our first principles calculations were performed using CASTEP code [22], a plane-wave implementation of density functional theory (DFT). The exchange-correlation (XC) contribution to the DFT energy was calculated using the GGA/PBE functional [23]. For comparison purposes, we also calculated the structure using LDA/CAPZ functional [24]. Convergence criteria employed for both the electronic self-consistent relaxation and the ionic relaxation were set to 10 eV and 0.006 eV/Å for energy and force, respectively.

**RESULTS AND DISCUSSIONS**

In Figure 2(a) we present the stress-strain curves for novamene under tensile deformation along x, y, and z-directions. Our results indicate a significant anisotropic behavior, quite distinct from the ones reported to graphene and diamond [25-30]. The abrupt drop of the stress values (ultimate strength) without a clear plastic region also indicates a brittle material. It should be stressed that although the x and y stress-strain curves are similar (in-plane strain), their deformation mechanisms (involving pentagons and/or hexagons) are different (inset of Figure 2(a)), which results in a slightly different ultimate strength (95.6 and 98.7 to x and y-directions, respectively). This chirality-dependent deformation mechanism is similar to what was reported for silicene [31] and chalcogenides [32].

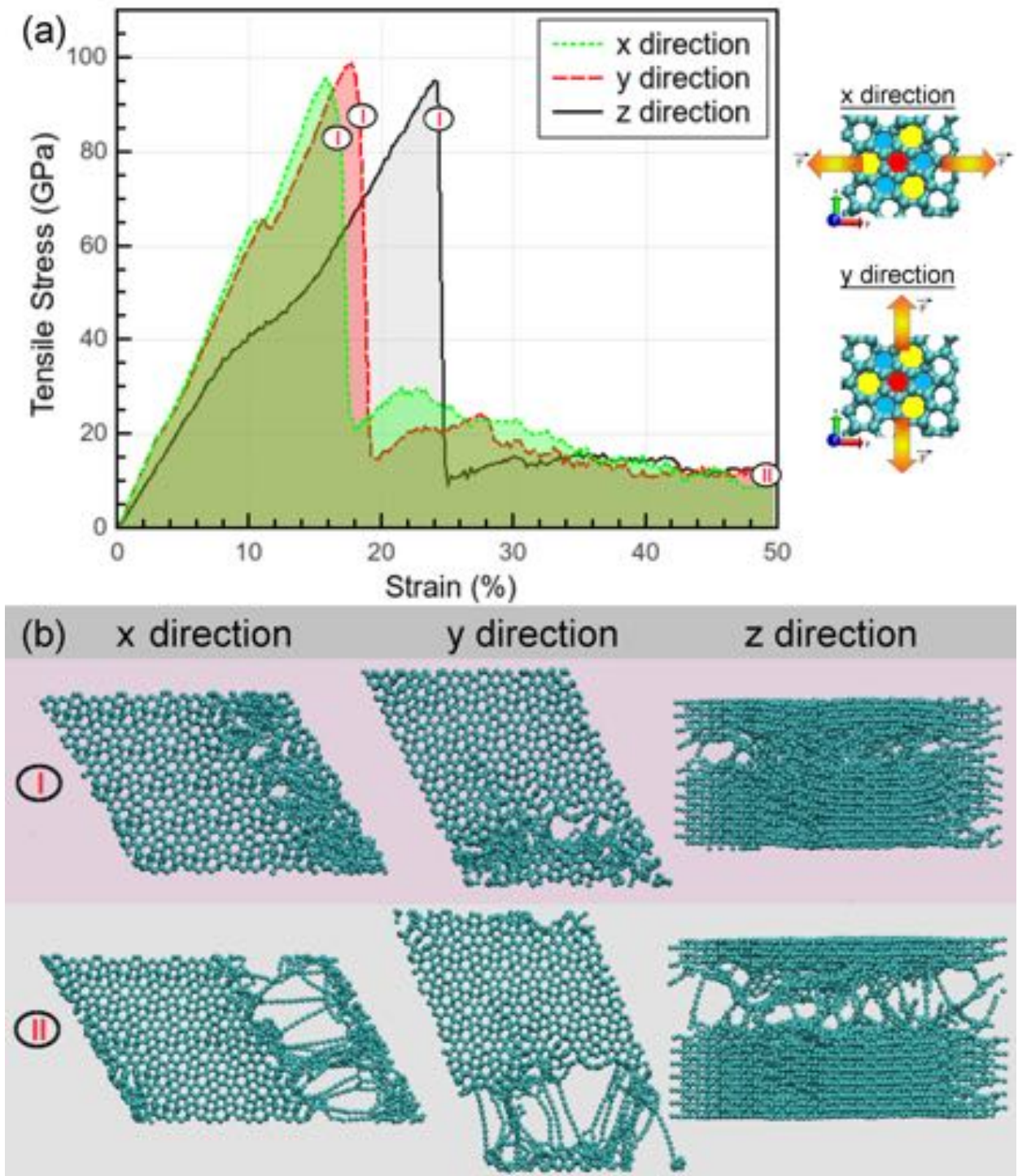

**Figure 2**. (a) Stress-strain curves for tensile deformation along x, y, and z novamene directions; the inset highlights that x and y deformations are not equivalent; (b) MD snapshots of two deformation stages for each tensile direction: (I) structure just after the structural failure/fracture and; (II) structure at 50% strain.

For the z-direction (inter-plane strain), the novamene deformation is more pronounced and reaches the ultimate strain at a higher value than the corresponding ones for x and y directions (24.5% in comparison to 15.5% and 17.5%, respectively). Some representative MD snapshots are presented in Figure 2(b). It is possible to observe that despite the brittle behavior, multiple linear carbon chains (LACs) are formed. LACs are common in fracture of carbon-based nanostructures and have been theoretically predicted and experimentally observed [33].

In Figure 3(a) we present the percentage amount of pure $sp^2$, $sp^3$, and broken bonds during the tensile elongation along the x-direction. Some representative MD snapshots are shown in Figure 3(b). As we can see, when novamene is stretched along the x-direction, the number of $sp^2$ and $sp^3$ carbon bonds decreases, while the number of broken bonds increases, as expected. The deformation behavior is almost linear up to ~10%, when it is observed a slight drop in the stress/strain curve (Figure 2(a)). A detailed analysis of this region (between 10.5% and 10.7%) showed that the number of $sp^2$, $sp^3$, and broken bonds remain practically constant (see inset of Figure 3(a)), which is suggestive that the novamene undergoes structural rearrangements, without the formation and/or breaking of covalent bonds. After this stage, the stress increases as well as the number of newly formed $sp^2$ bonds (from the heptagons), up to reaching a plateau that remains almost constant up to reaching the structural failure. After fracture, there is still a partial 'healing' with the formation of new chemical bonds, mainly in $sp^2$ hybridization form. Such results are corroborated by g(r) presented in the Figure S1, in supplementary material.

From the MD snapshots (Figure 3(b)), we can see that bond breaking occurs along the applied tensile direction, as expected (highlighted in red in Figure 3(b)). The novamene fracture produces multiple LACs, as evidenced in Figure 3(b) for strain at 48%. This formation suggests that some of the broken bonds can also form sp-$sp^2$, sp-$sp^3$, or sp-sp bonds. A similar behavior was observed for y-direction (see Figure S2 and S3 in SM).

For the z-direction the process is significantly different (Figure 3(c)). 10% strain is enough to increase the number of $sp^2$ bonds in relation to $sp^3$ ones (see also the corresponding g(r) - Figure S4). This $sp^2$ to $sp^3$ conversion keeps the number of broken bonds constant after increasing almost linearly. This conversion can continue up to the fracture limit (24.5%). The representative MD snapshots in Figure 3(d) show that the number of $sp^2$ bonds during the tensile elongation is closed related to the number "switching" carbon atoms along the z-direction. In the absence of applied strain, the "switching" carbon atoms are connected by $sp^3$ bonds. When the strain reaches values around 11%, the bonds related to the "switching atoms" become highly stretched and eventually breaks. At this moment, these bonds change to $sp^2$-type, as seen in the snapshot for 16% of strain (Figure 3(d)). As as consequence, the number of $sp^3$ bonds decreases, while the $sp^2$ ones increases. After fracture, multiple LACs are also present (see the snapshot for 48% of strain in Figure 3(d)). As we can see from Figure 3, in contrast to graphene, diamond and carbon nanotubes, novamene presents an extensive structural reconstructions (self-healing effect) with the formation of new covalent bonds ($sp^2$ and $sp^3$), not always from LACs.

In Figure 4 we present MD representative snapshots of the tensile procedure with atoms colored accordingly to von Mises stress values. The complete deformation processes are presented in the supplementary videos (Videos S1, S2, and S3). As can be seen from Figure 4, for x and y-directions, the stress per atom is initially well-distributed and concentrated mainly on the carbon atoms belonging to non-planar heptagons and hexagons. The fracture/crack propagation originated from these regions and the atoms composing the LACs came from broken heptagons and pentagons rings (see Videos S1 and S2).

In the case of z-direction, the stress is initially concentrated on the carbon atoms in $sp^3$ hybridization (including the "switching" carbon atoms), perpendicular to the tensile direction. Those regions will be the ones where the bonds will be broken. As the stress becomes more accumulated in the $sp^3$ bonds, it will deform the heptagons, and these deformations originate the fracture/crack propagation (see video S3). The remaining bonded atoms from the non-planar heptagons and hexagons also form the carbon lines as the novamene stretching continues. So, in general, the cracks in novamene come from the non-planar heptagons, hexagons and planar pentagons stretching during the tensile process.

In Table 1 we compare some mechanical properties calculated to novamene with other carbon nanostructures. The ultimate strength and strain, Young's modulus, and Poisson's ratio for graphene (armchair), carbon nanotube (12,0) and cubic diamond calculated using the same ReaxFF force field [20] are displayed in Table 1.

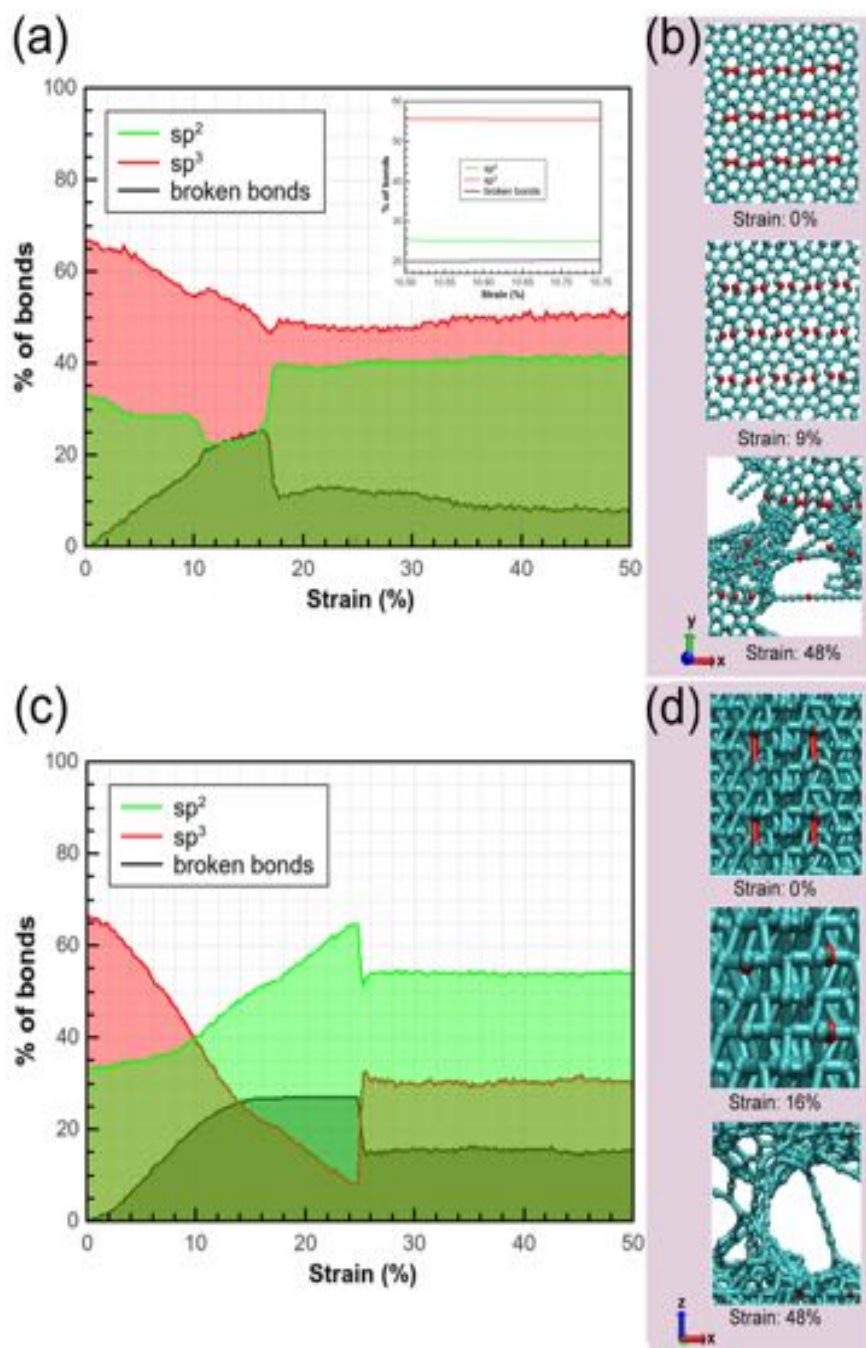

**Figure 3**. Percentage of the number of $sp^2$, $sp^3$, and broken bonds during the tensile testing for (a) x-direction with some; (b) representative MD snapshots for some strain levels, and; (c) z-direction with; (d) representative MD snapshots for some strain levels. The atoms highlighted in red indicate some broken bonds occurring during the tensile testing.

We also present the experimental data obtained from references [25-30, 34]. Our results show that novamene has an ultimate strength smaller than the other carbon allotropes. Regarding the ultimate strain, novamene along the z-direction has the highest one, which means it can stand larger deformations before fracture in comparison to the other listed carbon allotropes.

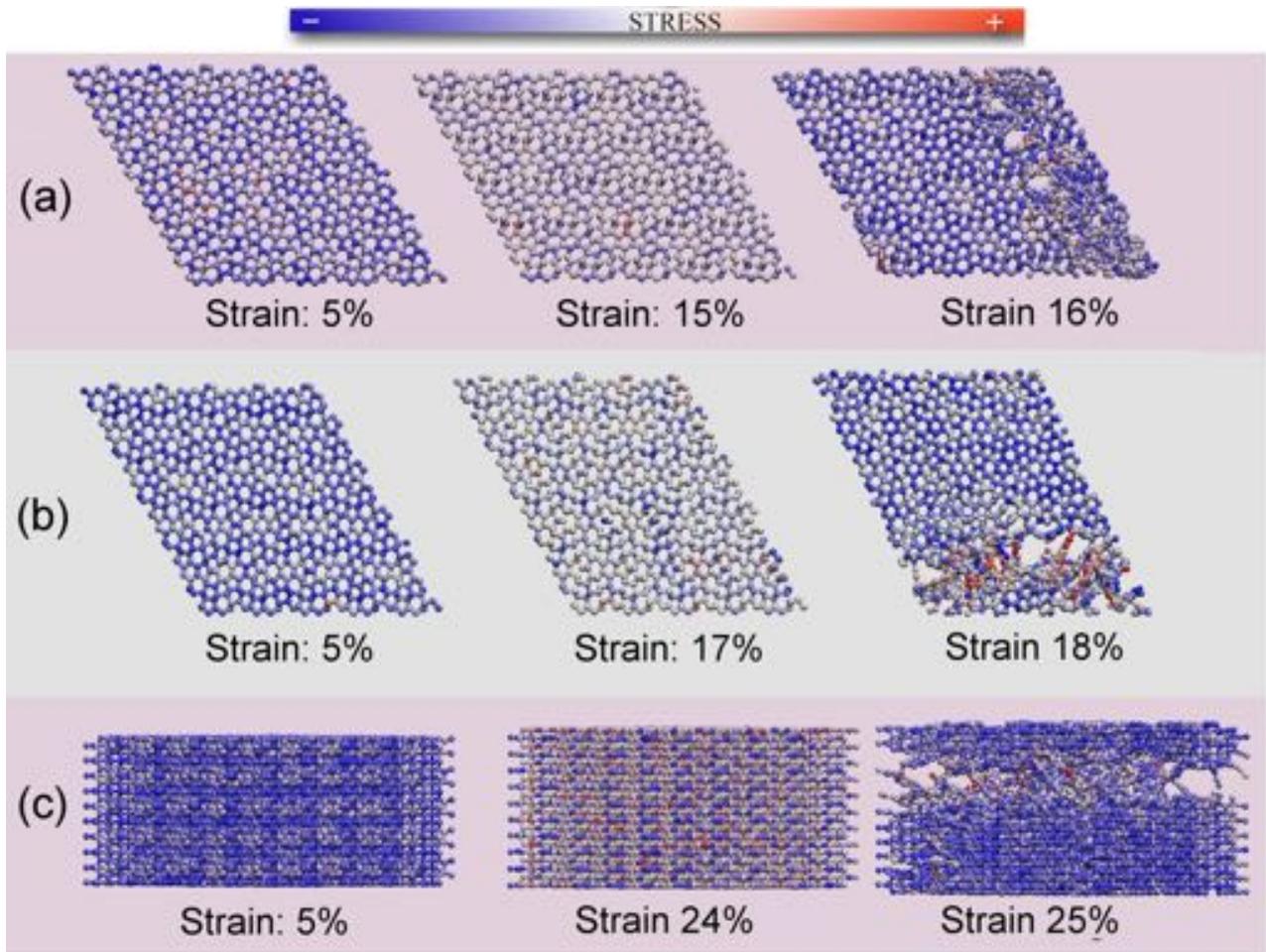

**Figure 4**. Stress accumulation during the tensile elongation along: (a) x, (b) y, and; (c) z-directions for some selected strain values: middle of the first linear regime and stages before and after the novamene mechanical failure/fracture. The colors represent the stress levels (from low (blue) to high (red) ones).

As mentioned before from the stress-strain analyses, novamene is an anisotropic material, thus it is expected different Young's modulus (E) values depending on deformation direction, as confirmed in Table 1. Young's modulus values are smaller than the others carbon allotropes considered here, which suggests that novamene is easier to deform, which is consistent with the Poisson's ratio valuers presented in Table 1 ($\sigma_{ij}$, where the **j** represents the tensile direction and **i** a perpendicular direction to the tensile one [34]) . Although Young's modulus and ultimate strength values are smaller than for the other structures listed in Table 1, they still outperform other materials, such as silicon, steel or titanium alloys [35].

**Table 1**. Mechanical properties obtained with ReaxFF force field for novamene (this work) and some selected carbon allotropes [20]; the available experimental data [25-30, 34] are presented into parentheses.

| Structure | Ultimate Strengh (GPa) | Ultimate Strain (%) | Young's Modulus (GPa) | Poisson's Ratio $\sigma_{ij}$ |
|---|---|---|---|---|
| Novamene (x) (this work) | 95.6 | 14.7 | 674.2 | $\sigma_{yx}$: = 0.60 $\sigma_{zx}$: = 0.04 |
| Novamene (y) (this work) | 98.7 | 16.3 | 659.7 | $\sigma_{xy}$: = 0.60 $\sigma_{zy}$: = 0.04 |
| Novamene (z) (this work) | 95.2 | 22.5 | 438.8 | $\sigma_{xz}$: =0.50 $\sigma_{yz}$: =0.50 |
| Diamond (z) | 148.7 (86-100) | 18.9 (12.0) | 1300.0 (1188.0) | $\sigma_{xz}$: 0.04 (0.07) $\sigma_{yz}$: 0.04 (0.04) |
| Graphene (armchair) | 149.6 (130.0) | 21.5 (25.0) | 1266.0 (1000.0) | 0.91 (~0.19) |
| Nanotube (12,0) | 111.3 (20-50) | 21.6 (1-6) | 1214.0 (320-1470) | ---- |

In order to obtain a better estimation of the elastic constant values, we also carried out DFT calculations using the functionals GGA/PBE and LDA/CAPZ, as discussed in the Materials and Methods section. We first optimized the novamente unit cell and then we calculated the energy band structure, shown in Figure 5. The calculations show the structure is a semiconductor and has an indirect band gap equals to 0.398 eV with GGA/PBE and 0.400 eV with LDA/CAPZ. These values are quite similar to the value 0.335 eV reported by L. A. Burchfield *et al.* [6]. We then calculated Young's modulus and Poisson's ratio values for the optimized structure. Although the DFT (see Table S1 of the supplementary material) and ReaxFF values are different, as expected, the general trends are consistent.

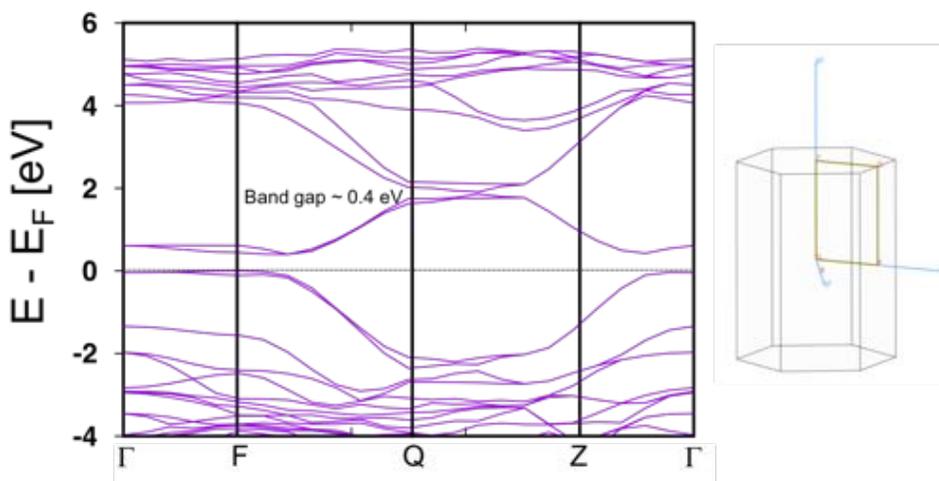

**Figure 5**. The energy band structure of novamene along a path accordingly to special points of the Brillouin zone shown in the right side of the figure.

**SUMMARY AND CONCLUSIONS**

We have investigated through fully atomistic reactive (ReaxFF potential) molecular dynamics and DFT (GGA/PBE and LDA/CAPZ functionals) simulations, the mechanical properties and fracture dynamics of a newly proposed carbon allotrope, named novamene. Novamene is a 3D structure composed of hexagonal diamond ($sp^3$ hybridization called lonsdaleite) and hexagonal carbon rings ($sp^2$ hybridization). In the present work we considered the single-ringed novamene structure.

Our results showed that novamene is an anisotropic structure with relation to tensile deformation, presenting similar behavior for x and y directions, but a distinct one for the z direction. Although novamente shares some mechanical features with other carbon allotropes, such as a brittle fracture behavior (diamond) and the formation of multiple carbon linear chains in the final fracture stages (graphene), it also exhibits distinct ones, such as, an extensive structural reconstructions (self-healing effect).

Novamene presents ultimate strength (~ 100 GPa) values lower than other carbon allotropes, but has the highest ultimate strain along the z-direction (~ 22.5%), which means it can stand larger deformations before fracture. Although the Young's modulus (~ 600 GPa) and ultimate strength values are smaller than for other carbon allotropes, they still outperform other materials, such as for example silicon, steel or titanium alloys.

With relation to the fracture dynamics, novamene is again anisotropic. While for the x and y directions the fracture/crack propagation originates from deformed heptagons and pentagons, for the z directions is mainly related to broken $sp^3$ bonds connecting structural planes. Another interesting feature is the formation of multiple and long carbon linear chains in the final fracture stages.

Further studies should be carried out to better characterize the novamene properties, as for example electrical and thermal transport, which could better determine its potential technological applications. Also, it will be interesting to investigate other types of novamene structures (double-ringed, triple-ringed, etc) in order to determine how the amount of initial $sp^2$ bonds (related to the number of carbon hexagons) into the hexagonal diamond structure will affect the mechanical properties of the resulting material. It was recently demonstrated for Schwarzites (another class of negative curvature carbon allotrope) [36], that the ratio of different structural rings (pentagons, hexagons, etc.) are of fundamental importance to determine the material response to tensile and/or compression deformation. We hope the present study will stimulate further studies along these lines.

**Acknowledgments**

We would like to thank the Brazilian agency FAPESP (Grants 2013/08293-7, 2014/24547-1 and 2016/18499-0) for financial support. Computational and financial support from the Center for Computational Engineering and Sciences at Unicamp through the FAPESP/CEPID Grant No. 2013/08293-7 is also acknowledged. Support from the Brazilian Agencies CNPq and CAPES is also acknowledged.

**References**

[1] Burchell TD. Carbon Materials for Advanced Technologies. 1st ed. Oxford: Elsevier Science, 1999.

[2] Messina G, Santangelo S. Carbon: The Future Material for Advanced Technology Applications. 1st ed. New York: Springer, 2006.

[3] Narayan J, Bhaumik A. Novel phase of carbon, ferromagnetism, and conversion into diamond, J. Appl. Phys. 2015;118:215303.

[4] He C, Sun L, Zhang C, Zhong J. Two viable three-dimensional carbon semiconductors with an entirely $sp^2$ configuration. Phys. Chem. Chem. Phys. 2013;15:680-684.

[5] Sheng X-L, Yan Q-B, Ye F, Zheng Q-R, Su G. T-Carbon: A novel narbon allotrope. Phys. Rev. Lett. 2011;106:155703.

[6] Burchfield LA, Fahim MA, Wittman RS, Delodovici F, Manini N. Novamene: A new class of carbon allotropes. Heliyon 2017;3:e00242.

[7] Nulakani NVR, Subramanian V. Cp-Graphyne: A low-energy graphyne polymorph with double distorted dirac points. ACS Omega 2017;2:6822-6830.

[8] Delodovici F, Manini N, Wittman RS, Choi DS, Fahim MA, Burchfield LA. Protomene: A new carbon allotrope. Carbon 2018;126:547-579.

[9] Wang X, Rong J, Song Y, Yu X, Zhan Z, Deng J. QPHT-graphene: A new two-dimensional metallic carbon allotrope. Phys. Lett. A 2017;381:2845-2849.

[10] Hu M, He J, Zhao Z, Strobel TA, Hu W, Yu D, Sun H, Liu L, Li Z, Ma M, Kono, Y. Compressed glassy carbon: An ultrastrong and elastic interpenetrating graphene network. Science advances 2017;3:e1603213.

[11] Vinod S, Tiwary CS, Autreto PAS, Taha-Tijerina J, Ozden S, Chipara AC, Vajtai R, Galvao DS, Narayanan TN, Ajayan PM. Low-density three-dimensional foam using self-reinforced hybrid two-dimensional atomic layers. Nat. Commun. 2014;5:4541.

[12] Frondel C, Marvin UB. Lonsdaleite, a hexagonal polymorph of diamond. Nature 1967;214:587–589.

[13] Bundy FP, Kasper JS. Hexagonal Diamond—A new form of carbon. J. Chem. Phys. 1967;46:3437.

[14] Bruice PY. Organic Chemistry. 3rd ed. New Jersey: Pearson, 2001.

[18] Frenkel D, Smit B. Understanding molecular simulation: From algorithms to applications. 2nd ed. Academic Press: San Diego, 2001.

[19] Zang A, Stephansson O. Stress field of the earth's crust. 1st ed. Springer: Houten, 2009.

[20] Jensen BD, Wise KE, Odegard GM. The effect of time step, thermostat, and strain rate on ReaxFF simulations of mechanical failure in diamond, graphene, and carbon nanotube. J. Comput. Chem 2015;36:1587-1596.

[21] Jensen BD, Wise KE, Odegard GM. Simulation of the Elastic and Ultimate Tensile Properties of Diamond, Graphene, Carbon Nanotubes, and Amorphous Carbon Using a Revised ReaxFF Parametrization. J. Phys. Chem. A 2015;119:9710−9721.

[22] Clark SJ, Segall MD, Pickard CJ, Hasnip PJ, Probert MIJ, Refson K, Payne MC. First principles methods using CASTEP. Kristallogr. Cryst. Mater. 2005;220:567-570.

[23] Perdew JP, Burke K, Ernzerhof M. Generalized gradient approximation made simple. Phys. Rev. Lett.,1996; 77:3865-3868.

[24] Perdew JP, Zunger A. Self-interaction correction to density-functional approximations for many-electron systems. Phys. Rev. B 1981;23:5048.

[25] Schneider T, Stoll E. Molecular-dynamics study of a three-dimensional one-component model for distortive phase transitions. Phys. Rev. B 1978;17:1302.

[26] McSkimin HJ, Bond WL. Elastic moduli of diamond. Phys. Rev. 1957;105:116.

[27] Field JE, Pickles CSJ. Strength, fracture and friction properties of diamond. Diamond Relat. Mater. 1996;5:625-634.

[28] Telling RH, Pickard CJ, Payne MC, Field JE. Theoretical strength and cleavage of diamond. Phys. Rev. Lett. 2000;84:5160.

[29] Lee C, Wei X, Kysar JW, Hone J. Measurement of the elastic properties and intrinsic strength of monolayer graphene. Science 2008;321:385-388.

[30] Liu F, Ming P, Li J. Ab initio calculation of ideal strength and phonon instability of graphene under tension. Phys. Rev. B 2007;76:064120.

[31] Botari T, Perim E, Autreto PAS, van Duin ACT, Paupitz R, Galvao DS. Mechanical properties and fracture dynamics of silicene membranes. Phys. Chem. Chem. Phys., 2014;16:19417-19423.

[32] Manimunda P, Nakanishi Y, Jaques YM, Susarla S, Woellner CF, Bhowmick S, Asif SAS, Galvao DS, Tiwary CS, Ajayan PM. Nanoscale deformation and friction characteristics of atomically thin $WSe_2$ and heterostructure using nanoscratch and Raman spectroscopy. 2D Materials 2017;4:045005.

[33] Andrade NF, Aguiar AL, Kim YA, Endo M, Freire PTC, Brunetto G, Galvao DS, Dresselhaus MS, Souza Filho AG. Linear carbon chains under high-pressure conditions. J. Phys. Chem. C 2015;119:10669-10676.

[34] Shackelford, JF. Introduction to materials science for engineers. 6th ed. Pearson: London, 2015.

[35] Yu MF, Files BS, Arepalli S, Ruoff RS. Tensile loading of ropes of single wall carbon nanotubes and their mechanical properties. Phys. Rev. Lett. 2000;84:5552.

[36] Sajadi SM, Owuor PS, Schara S, Woellner CF, Rodrigues V, Vajtai R, Lou J, Galvao DS, Tiwary CS, Ajayan PM. Multiscale Geometric Design Principles Applied to 3D Printed Schwarzites. Adv. Mater. 2018;30:1704820.